\documentclass[aps,prb,reprint,groupedaddress]{revtex4-1}

\usepackage{amssymb}
\usepackage{amsmath}
\usepackage{graphicx}
\usepackage{epsfig}

\begin{document}

\title{Tuning in-plane FFLO state in superconductor-ferromagnet-normal metal hybrid structure by magnetic field or current}

\author{P. M. Marychev}
\email[Corresponding author: ]{marychevpm@ipmras.ru}
\author{D. Yu. Vodolazov}
\affiliation{Institute for Physics of Microstructures, Russian
Academy of Sciences, Nizhny Novgorod, 603950 Russia}

\date{\today}

\begin{abstract}

Temperature induced transition of thin
superconductor-ferromagnet-normal (S/F/N) metal hybrid structure
to in-plane Fulde-Ferrel-Larkin-Ovchinnikov (FFLO) state is
accompanied by vanishing of effective inverse magnetic field
penetration depth $\Lambda^{-1}$ (Phys. Rev. Lett. {\bf 121},
077002 (2018)). Here we show that $\Lambda^{-1}$ goes to zero only
in limit of zero magnetic field $H \to 0$ and at any finite
parallel $H$ or in-plane current $I$ it is finite and positive in
FFLO state which implies diamagnetic response. We demonstrate that
$\Lambda^{-1}$ has a nonmonotonic dependence on $H$ and $I$ not
only in the parameter range corresponding to the FFLO phase domain
but also in its vicinity. We find that for S/F/N/F/S structures
with certain thicknesses of F layers there is temperature, current
and magnetic field driven transition to and out of FFLO phase with
a simultaneous jump of $\Lambda^{-1}$.
\end{abstract}

\maketitle

\section{Introduction}

In superconductor-ferromagnet (S/F) bilayers proximity induced
odd-frequency spin triplet superconducting component in F layer
gives {\it negative} contribution to square of inverse London
penetration depth $\lambda^{-2}$
[\onlinecite{Bergeret_2001,Asano_2011,Yokoyama_2011,Asano_2014,Alidoust_2014,Fominov_2015}],
which is a coefficient in relation between superconducting current
density and vector potential: $\textbf{j}=-c\textbf{A}/4\pi
\lambda^2$. At some parameters this contribution can exceed
positive contribution from singlet superconducting component in S
and F layers and makes effective inverse magnetic field
penetration depth $\Lambda^{-1}=\int_0^d \lambda^{-2}(x) dx$ ($d$
is a thickness of the bilayer) negative which implies paramagnetic
response of whole structure. In Ref. \cite{PRL-2012} it is argued
that the state with $\Lambda^{-1}<0$ is unstable and authors find
that the S/F bilayer transits to in-plane
Fulde-Ferrel-Larkin-Ovchinnikov (FFLO) state as $\Lambda^{-1} \to
+0$. In the recent work \cite{PRL-2018} it is predicted that such
an in-plane FFLO state can emerge at temperature much below the
critical one, it is characterized by unusual current-phase
relation and can be realized in S/F/N trilayer with realistic
parameters, where N is a low resistive normal metal (Au, Ag, Cu or
Al), S is a disordered superconductor with large residual
resistivity in the normal state (NbN, WSi, NbTiN etc.) and F is
ordinary ferromagnet (Fe, CuNi, etc.).

Motivated by these results and expected unusual electrodynamic
response of FFLO state in S/F/N trilayer with rather usual
parameters, easily realizable by modern experimental technique, we
theoretically study effect of parallel magnetic field $H$ and
in-plane current $I$ on FFLO state in S/F/N trilayer and S/F/N/F/S
symmetric pentalayer. We find that $\Lambda^{-1}=0$ only in the
limit $H, I\to 0$ and it is positive for any finite magnetic field
or current which means that S/F/N trilayer has a diamagnetic
response. Parallel magnetic field and in-plane current suppresses
proximity induced odd-frequency triplet superconductivity in F/N
layers and $\Lambda^{-1}$ increases in weak magnetic field
(current). The same effect exists for trilayer being close to FFLO
phase domain ($\Lambda^{-1} \neq 0$ at $H,I=0$) due to
contribution of triplet component to $\Lambda^{-1}$. In pentalayer
FFLO phase domain is smaller due to competition of FFLO state with
$\pi$ state (well-known for S/F/S trilayers \cite{RMP-2005}) but
there are temperature, magnetic field and current driven
transitions from $\pi$ to the FFLO state with considerable change
of $\Lambda^{-1}$.

The structure of the paper is following. In section II we present
our theoretical model. In section III we show our results on
effect of parallel magnetic field and in-plane current on
$\Lambda^{-1}$ in S/F/N trilayer being in FFLO state or in state
with large contribution of odd-frequency triplet component to
$\Lambda^{-1}$. In section IV we consider different types of the
$\pi\to$ FFLO transitions in S/F/N/F/S structures and their
influence on the screening properties. Section V contains a brief
summary.

\section{Model}

\begin{figure}[hbt]
 \begin{center}
\includegraphics[width=1.0\linewidth]{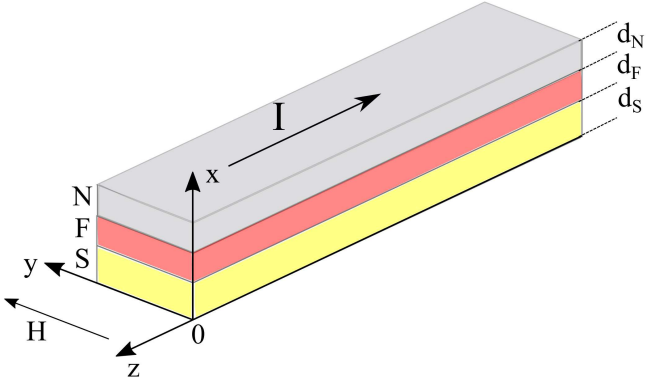}
 \caption{\label{Fig:Sys}
The schematic representation of the S/F/N structure under
consideration with transport current $I$ or placed in parallel
magnetic field $H$.}
 \end{center}
\end{figure}

To study the superconducting properties of S/F/N and S/F/N/F/S
structures we use the one-dimensional Usadel equation
\cite{Usadel} for normal $g$ and anomalous $f$ quasi-classical
Green functions. With standard angle parametrization $g=cos\Theta$
and $f=sin\Theta\exp(i\varphi)$ the Usadel equations in different
layers can be written as

\begin{equation}
 \label{usadel-s}
\frac{\hbar D_S}{2}\frac{\partial^2\Theta_S}{\partial
x^2}-\left(\hbar\omega_n+\frac{D_S}{2\hbar}q^2\cos
\Theta_S\right)\sin\Theta_S+\Delta\cos\Theta_S=0,
 \end{equation}
 \begin{equation}
  \label{usadel-f}
\frac{\hbar D_F}{2}\frac{\partial^2\Theta_F}{\partial
x^2}-\left((\hbar\omega_n+ih)+\frac{D_F}{2\hbar}q^2\cos
\Theta_F\right)\sin\Theta_F=0,
 \end{equation}
 \begin{equation}
  \label{usadel-n}
  \frac{\hbar D_N}{2}\frac{\partial^2\Theta_N}{\partial
x^2}-\left(\hbar\omega_n+\frac{D_N}{2\hbar}q^2\cos
\Theta_N\right)\sin\Theta_N=0,
\end{equation}
where subscripts S, F and N refer to superconducting,
ferromagnetic and normal layers, respectively. Here $D$ is the
diffusion coefficient for corresponding layer, $h$ is the exchange
field in F layer, $\hbar \omega_n = \pi k_BT(2n+1)$ are the
Matsubara frequencies ($n$ is an integer number), $q=\nabla\varphi
+ 2\pi\,{\bf A}/\Phi_0$ is the quantity that is proportional to
supervelocity $v_s=\hbar q/m$ directed in z direction (see Fig.
\ref{Fig:Sys}), $\varphi$ is the phase of the order parameter,
${\bf A}$ is the vector potential, $\Phi_0=\pi\hbar c/|e|$ is the
magnetic flux quantum. The $x$--axis is oriented perpendicular to
the surface of S layer accordingly to Fig. \ref{Fig:Sys}. $\Delta$
is the superconducting order parameter, which satisfies to the
self-consistency equation

\begin{equation}
\label{self-cons}
\Delta \ln\left(\frac{T}{T_{c0}}\right)=2\pi k_B
T\sum_{\omega_n
>0} \Re\left(\sin\Theta_S - \frac{\Delta}{\hbar\omega_n}\right),
\end{equation}
where $T_{c0}$ is the critical temperature of single S layer
(film) in the absence of magnetic field. These equations are
supplemented by the Kupriyanov-Lukichev boundary conditions
between layers \cite{JETP-1988}
  \begin{eqnarray}
  \nonumber
    \left.D_S\frac{d\Theta_S}{dx}\right|_{x=d_S-0}=\left.D_f\frac{d\Theta_F}{dx}\right|_{x=d_S+0},
    \\
    \label{boundary}
    \left.D_F\frac{d\Theta_F}{dx}\right|_{x=d_S+d_F-0}=\left.D_N\frac{d\Theta_N}{dx}\right|_{x=d_S+d_F+0}
    \end{eqnarray}
For the sake of simplicity we assume that the barrier between
layers does not exist and thereupon $\Theta$ is continuous
function of x. For interfaces with vacuum we use the boundary
condition $d\Theta/dx=0$. In the $\pi$-state we add the condition
$\Theta=0$ in the middle of the pentalayer structure.

We assume that the thickness of whole structure is much smaller
than the London penetration depth $\lambda$ of the single S layer
and neglect the effect of screening on the vector potential and
magnetic field. For chosen direction of the applied magnetic field
(see Fig. 1) we use the following vector potential: ${\bf
A}=(0,0,-Hx)$ in case of trilayer and ${\bf
A}=(0,0,-H(x-d_S-d_F-d_N/2))$ in case of pentalayer.

To calculate the supercurrent density we use the following
expression
\begin{equation}
 \label{current}
j=\frac{2\pi k_BT}{e\rho}q\sum_{\omega_n > 0}\Re(\sin^2\Theta),
\end{equation}
where $\rho$ is the residual resistivity of the corresponding
layer. From Eq. (6) and London relation $j=-cA/4\pi \lambda^{2}$,
one can find expression for square of inverse London penetration
depth
\begin{equation}
\label{lambda-local} \frac{1}{\lambda^2(x)}=\frac{16
\pi^2k_BT}{\hbar c^2\rho}\sum_{\omega_n > 0} \Re(\sin^2\Theta),
\end{equation}
and for the inverse effective penetration depth
 \begin{equation}
 \label{lambda-eff}
    \Lambda^{-1}=\int\limits_{0}^{d} \frac{dx}{\lambda^2(x)},
 \end{equation}
where the total thickness $d=d_S+d_F+d_N$ for the S/F/N and
$d=2d_S+2d_F+d_N$ for the S/F/N/F/S structures. In case of thin S
film $\Lambda$ coincides with Pearl penetration depth
\cite{Pearl}.

Because we neglect variation of $H$ due to screening we simply use
the Helmholtz free energy per unit of square
\begin{multline}
F_{H}=\pi N(0)k_BT\sum_{\omega_n \geq 0}\int \Re\{\hbar
D[(\nabla \Theta)^2+\sin^2\Theta(q/\hbar)^2]
\\
\label{free-energy}
 -4(\hbar\omega_n+ih)(\cos\Theta-1)-2\Delta \sin\Theta \}dx.
\end{multline}

In numerical calculations we use the dimensionless units. The
magnitude of the order parameter is normalized in units of
$k_BT_{c0}$, length is in units of $\xi_c=\sqrt{\hbar
D_S/k_BT_{c0}}$, the free energy per unit of square is in units of
$F_0=N(0)(k_BT_{c0})^2 \xi_c$. The magnetic field is measured in
units of $H_0=\Phi_0/2\pi\xi_c^2$, the effective penetration depth
is in units of $\Lambda=\lambda_0^2/d_S$, where $\lambda_0$ is the
London penetration depth of the single S layer at zero
temperature.

To find the effective penetration depth $\Lambda^{-1}$, we
numerically solve equations \ref{usadel-s}~--~\ref{self-cons},
using Kupriyanov-Lukichev boundary conditions \ref{boundary}. In
calculations we assume that the density of states on the Fermi
level $N(0)$ is the same for all layers and, therefore, the ratio
of resistivities is inversely proportional to the ratio of
corresponding diffusion coefficients. To reduce the number of free
parameters we also assume that the resistivity of S layer and F
layers are equal, i.e. $\rho_S/\rho_F=1$, which roughly
corresponds to parameters of real S and F films. Because formation
of FFLO state in the S/F/N structure needs the large ratio of
resistivities between N layer and S layers, we use
$\rho_S/\rho_N=150$ in our calculations, which is close to the
parameters of real materials \cite{PRL-2018}. For example for pair
NbN/Al the ratio $\rho_S/\rho_N$ could be as large as 400 (Ref.
\onlinecite{Vodolazov_SUST}) while for pair NbN/CuNi
$\rho_S/\rho_F \sim 1.5$ (Ref. \onlinecite{Yamashita}). The
exchange field of the ferromagnet $h$ is assumed to be of the
order of the Curie temperature $T_{curie}$ (for example in CuNi
\cite{Yamashita} $h\sim 13k_BT_{c0}$).

In-plane FFLO state could be realized as FF-like state (in this
case $f(z)\sim exp(iq_0z)$) or as LO-like state (in this case
$f(z)\sim cos(q_0z)$ near $T^{FFLO}$). In addition to 1D
calculations we also numerically solved 2D Usadel equation (in x
and z directions in Fig. \ref{Fig:Sys}) with following boundary
conditions along z direction: $f(z=0)=f(z=\pi/q_0)=0$ and found
that such a LO-like state has an energy (per unit of volume)
larger than FF-like state at any $q_0=\partial \varphi/\partial
z$. Therefore throughout our paper under FFLO state we assume
FF-like state $f(z)\sim exp(iq_0z)$ and solve 1D problem in x
direction.

\section{S/F/N trilayer}

Let us first consider the S/F/N trilayer. As it is shown in
\cite{PRL-2018}, in the case $\rho_N \ll \rho_S$ there is a range
of parameters when in-plane FFLO phase appears below the certain
critical temperature $T^{FFLO}<T_c$. In the FFLO phase the
effective penetration depth $\Lambda^{-1}=0$ as $H \to 0$, which
signals about vanishing of magnetic response at $T\leq T^{FFLO}$.
Here we calculate effect of finite $H$ and $I$ on $\Lambda^{-1}$.

\begin{figure}[hbt]
 \begin{center}
\includegraphics[width=1.0\linewidth]{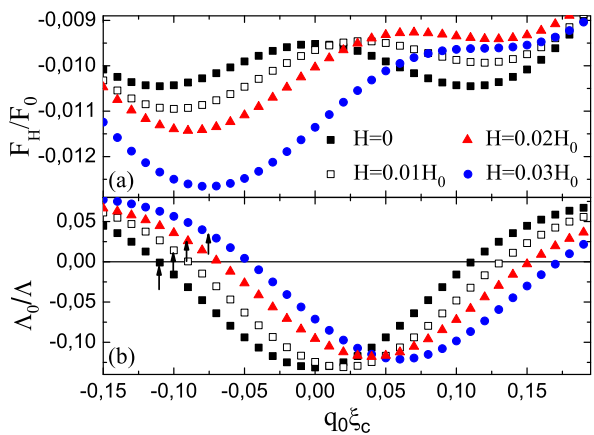}
 \caption{\label{Fig:F(q)}
Dependence of the free energy $F_H$ (a) and inverse effective
penetration depth $\Lambda^{-1}$ (b) on $q_0$ for the S/F/N
trilayer being in in-plane FFLO state at different values of the
parallel magnetic field. The arrows indicate the values of
$\Lambda^{-1}$ corresponding to the left minimum of the free
energy. We use the following parameters of the system:
$h=5k_BT_{c0}$, $d_S=1.1\xi_c$, $d_F=0.5\xi_c$, $d_N=\xi_c$ and
$T=0.3T_{c0}$. }
 \end{center}
\end{figure}

In Fig. \ref{Fig:F(q)}(a) we show dependence $F_H(q_0)$ for the
trilayer below the FFLO transition temperature $T^{FFLO}$. One can
see that in the absence of the external field two states with $q_0
\neq 0$ have a minimal energy. Both states correspond to
$\Lambda^{-1}=0=\partial F_H/\partial q_0$ (Fig.
\ref{Fig:F(q)}(b)). The parallel magnetic field $H$ breaks the
symmetry $F_H(q_0)$ and leads to the increase and furthermore
disappearance of one of the energy minimums (right one in Fig.
\ref{Fig:F(q)}(a)). Corresponding to this minimum state has
negative value of $\Lambda^{-1}$ at $H > 0$ and according to the
arguments suggested in Ref. \onlinecite{PRL-2012} should be
considered as an unstable one. Indeed, one can show that the term
corresponding to contribution of kinetic energy to $F_H$ is
proportional to $\Lambda^{-1}q^2$. When $\Lambda^{-1}<0$ it is
energetically favorable to have nonzero supervelocity $\sim q$ (if
it was zero) or increase it (if it was finite) which makes such a
state with negative $\Lambda^{-1}$ unstable. To see how this
instability evolves in time and what is the finite state one
should solve 3D problem in our case (taking into account $q\neq 0$
in all directions) and it is out of scope of the present research.
Further we consider only the state with $\Lambda^{-1}\geq 0$,
corresponding to the left minimum of $F_H(q_0)$ in Fig.
\ref{Fig:F(q)}(a).

\begin{figure}[hbt]
 \begin{center}
\includegraphics[width=1.05\linewidth]{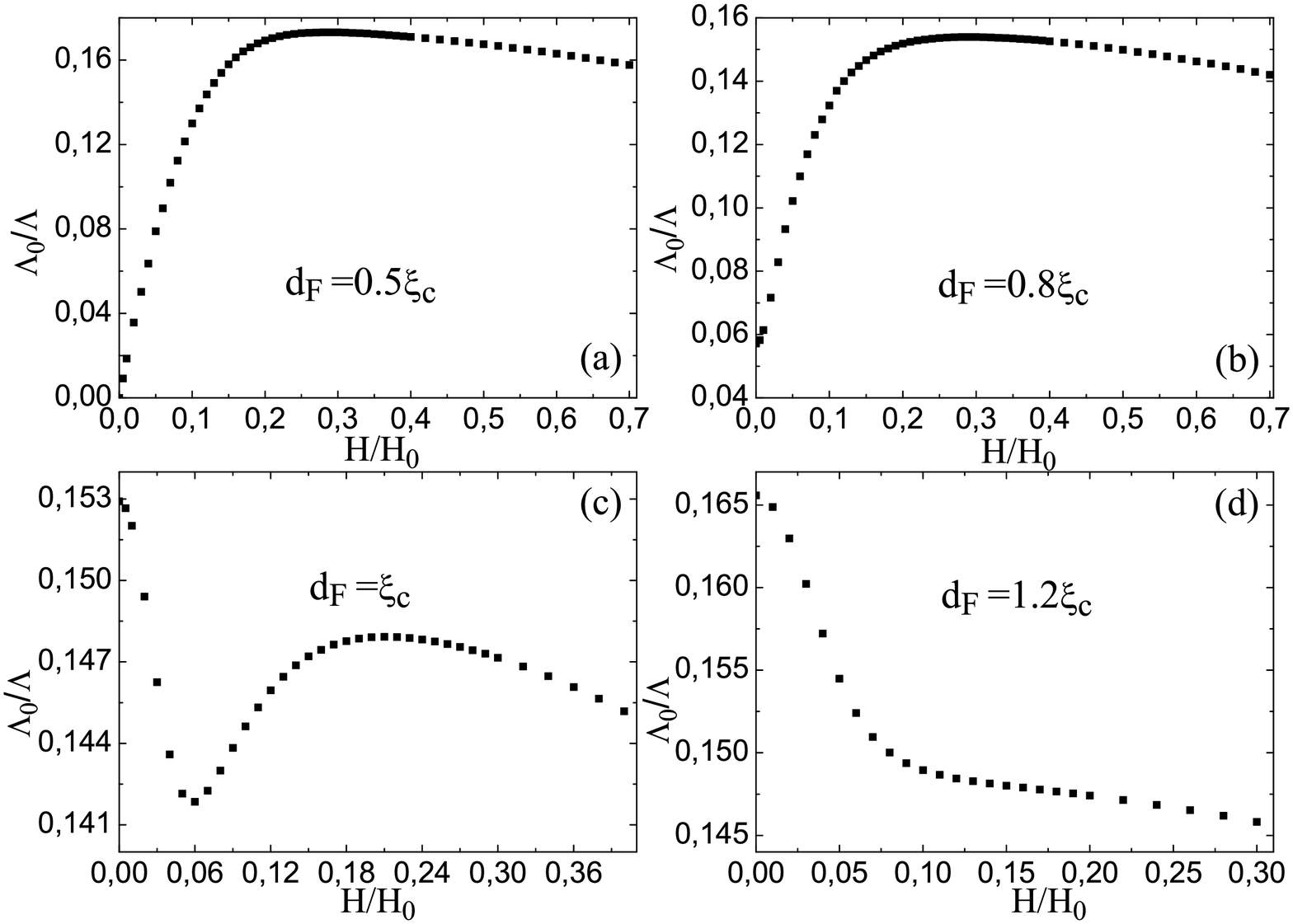}
 \caption{\label{Fig:Lambda}
Dependence of the inverse effective penetration depth of the
magnetic field $\Lambda^{-1}$ in the S/F/N trilayer on the
parallel magnetic field $H$ at different thicknesses of F layer
$d_F$: (a) $0.5\xi_c$ (FFLO state); (b) $0.8\xi_c$; (c) $\xi_c$;
(d) $1.2\xi_c$. The other parameters of the trilayer are
following: $h=5k_BT_{c0}$, $d_S=1.1\xi_c$, $d_N=\xi_c$ and
$T=0.2T_{c0}$.}
 \end{center}
\end{figure}

\begin{figure}[hbt]
 \begin{center}
\includegraphics[width=1.0\linewidth]{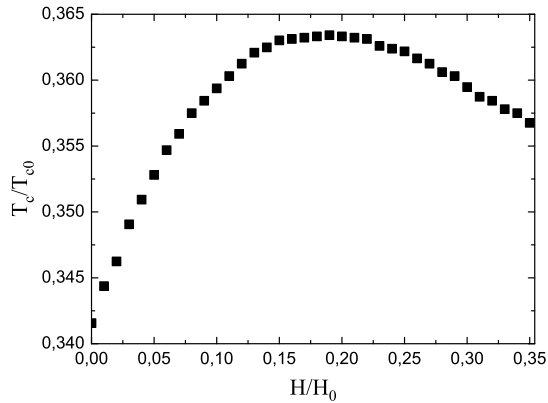}
 \caption{\label{Fig:tc(h)}
Dependence of the critical temperature of S/F/N trilayer on the
parallel magnetic field $H$. The parameters of the system are
following: $h=25k_BT_{c0}$, $d_S=1.2\xi_c$, $d_F=0.16\xi_c$,
$d_N=\xi_c$. Using smaller thickness of S layer one can obtain
larger relative change of $T_c$ but $T_c$ itself goes to lower
temperatures. The similar result could be obtained for
$h=5k_BT_{c0}$ too.}
 \end{center}
\end{figure}

The field dependence of $\Lambda^{-1}\geq 0$ is present in Fig.
\ref{Fig:Lambda}(a). One can see that $\Lambda^{-1}$
nonmonotonically changes with a field. Increase of $\Lambda^{-1}$
at relatively weak magnetic field is connected with two effects.
The first one is the suppression of the superconducting
correlations (including triplet one) in N layer by magnetic field
and we find that it gives main contribution to increase of
$\Lambda^{-1}$. Besides that there is slight enhancement of of
singlet superconductivity in S layer, because weak magnetic field
decreases supervelocity $\sim q = q_0+2\pi A/\Phi_0$ in S layer,
and it also provides enhancement of $\Lambda^{-1}$. The second
effect is responsible for enhancement of $T_c^{FFLO}$ (see Fig.
\ref{Fig:tc(h)}) by applied field - earlier this effect was
predicted for S/F bilayer being in FFLO state in Ref.
\onlinecite{JETPL-2014}. Note that the found enhancement of $T_c$
is rather small for S/F/N trilayer with realistic parameters.

Sufficiently large magnetic field destroys proximity-induced
superconductivity in F/N layers and $\Lambda^{-1}$ reaches the
maximum value - see Fig.\ref{Fig:Lambda}(a). The following
decrease of $\Lambda^{-1}$ is explained by gradual increase of $q
\sim A$ in S layer and gradual suppression of $|\Delta|$ as in
usual S film. These results show that $\Lambda^{-1}$ is finite and
positive at any finite $H$ for S/F/N trilayer being in FFLO state.
One may also conclude that the magnetic response is diamagnetic
and nonlinear even at $H\to 0$ because $\Lambda^{-1}$ changes from
zero up to the finite value.

Even if $\Lambda^{-1}$ is positive at $H=0$ and trilayer is not in
the FFLO state the dependence $\Lambda^{-1}(H)$ may be
nonmonotonic due to contribution of triplet component to
$\Lambda^{-1}$. In Fig. \ref{Fig:Lambda}(b,c,d) we demonstrate it
by varying the thickness of F layer and keeping over parameters of
trilayer constant. With increasing $d_F$ contribution of triplet
component to $\Lambda^{-1}$ decreases, but it stays finite. Small
increase of $d_F$ (see Fig.\ref{Fig:Lambda}(b)) drives the system
from FFLO state but due to considerable contribution of triplet
component dependence $\Lambda^{-1}(H)$ resembles one shown in Fig.
\ref{Fig:Lambda}(a). In Ref. \onlinecite{PRL-2018} somewhat
related effect is found for dependence $\Lambda^{-1}(T)$ in the
vicinity of the FFLO domain. Consequently, the increase of
$\Lambda^{-1}$ with magnetic field can serve as a precursor of the
FFLO state as the increase of $\Lambda^{-1}$ with increasing of
temperature.\cite{PRL-2018}

At larger $d_F$, i.e. with getting further from the FFLO domain,
$\Lambda^{-1} (H=0)$ increases and starting from some value of
$d_F$ ($\approx 2\sqrt{\hbar D_F/h}$) the inverse penetration
depth $\Lambda^{-1}$ decreases in a weak magnetic field (see Fig.
\ref{Fig:Lambda}(c)). Our calculations show that the effect is
connected with faster decay of singlet component than the triplet
one in N layer at weak magnetic field. At the field larger some
value (it roughly corresponds to the minimum in dependence
$\Lambda^{-1}(H)$ shown in Fig.\ref{Fig:Lambda}(c)) the proximity
induced superconductivity is getting suppressed stronger and
$\Lambda^{-1}$ increases as in \ref{Fig:Lambda}(a,b) and the
dependence $\Lambda^{-1}(H)$ has both minimum and maximum. Further
increase of $d_F$ (see \ref{Fig:Lambda}(d)) leads to the monotonic
decrease of $\Lambda^{-1}$ in magnetic field (triplet component
gives small contribution to $\Lambda^{-1}$) and the influence of N
layers manifests itself in rapid vanishing of $\Lambda^{-1}$ at
relatively weak fields when superconductivity is terminated there.

Very similar results we obtain for the larger value of exchange
field ($h=25 k_BT_{c0}$) with the only difference that they occur
in much narrower range of $d_F$ with respect to $\xi_c$,
reflecting smaller value of characteristic decay length of
superconducting correlations in F layer $\xi_F\sim 1\sqrt{h}$
(results are not shown here). Qualitatively the same dependencies
$\Lambda^{-1}(H)$ (except the one with two extremum shown in Fig.
\ref{Fig:Lambda}(c)) could be found at fixed $d_F$ when one
increases temperature from $T<T^{FFLO}$ up to $T^{FFLO}<T<T_c$
when the trilayer is driven from in-plane FFLO to uniform state
but with still noticeable contribution of triplet
superconductivity to $\Lambda^{-1}$.

\begin{figure}[hbt]
 \begin{center}
\includegraphics[width=1.05\linewidth]{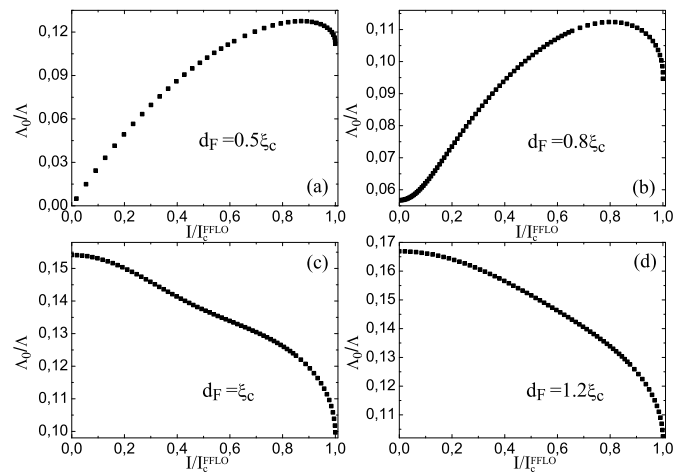}
 \caption{\label{Fig:Lambda-current}
Dependence of the inverse effective penetration depth
$\Lambda^{-1}$ in the S/F/N trilayer on the in-plane $I$ at
different thicknesses of F layer $d_F$: (a) $0.5\xi_c$ (FFLO
state); (b) $0.8\xi_c$; (c) $\xi_c$; (d) $1.2\xi_c$. Current is
expressed in units of the critical current of the FFLO-state
$I_c^{FFLO}$. The rest of parameters is the same as in
Fig.\ref{Fig:Lambda}.}
 \end{center}
\end{figure}

FFLO state in S/F/N trilayer could be tuned not only by parallel
magnetic field but in-plane current too. As in the case of
parallel magnetic field applied current breaks the symmetry $q_0
\to -q_0$ and in Fig. 5 we show dependence of $\Lambda^{-1}\geq 0$
on the in-plane current for the same parameters as in Fig. 3.
External current (supervelocity) suppresses stronger proximity
induced superconductivity in N layer than the superconductivity in
S layer (like in S/N bilayer \cite{Vodolazov_SUST}) and
$\Lambda^{-1}$ increases with the current for some $d_F$ (see
Figs. \ref{Fig:Lambda-current}(a,b)). Qualitatively, the results
shown in Fig. \ref{Fig:Lambda-current}(a) could be found using
modified Ginzburg-Landau equation \cite{Bouzdin} as it was done in
Ref. \cite{PRB-2017} where current-carrying FFLO state was
studied. This approach is much simpler than the used here Usadel
equations and allows to obtain analytical solution for current
states but it has two disadvantages: i) it cannot be used to study
states which are not in the FFLO phase domain and ii) it is
difficult to relate coefficients in Ginzburg-Landau functional
with microscopic parameters of S/F/N structure.

\section{$\pi\to$ FFLO transition in S/F/N/F/S structures}

Let us discuss now symmetric S/F/N/F/S pentalayer (it could be
imagined as doubled trilayer). Our interest to this system is
mainly connected with existence of $\pi$ state, corresponding to
the phase difference $\pi$ between outer S layers, together with 0
state (uniform or FFLO one) considered in previous section. We
restrict ourself by consideration of the uniform $\pi$-state,
because in the chosen parameter range modulated (FFLO) $\pi$-state
is not realized (note that in the recent work \cite{PSS-2017} such
a state is predicted for the S/F/S structure in certain range of
parameters).
\begin{figure}[hbt]
 \begin{center}
\includegraphics[width=1.0\linewidth]{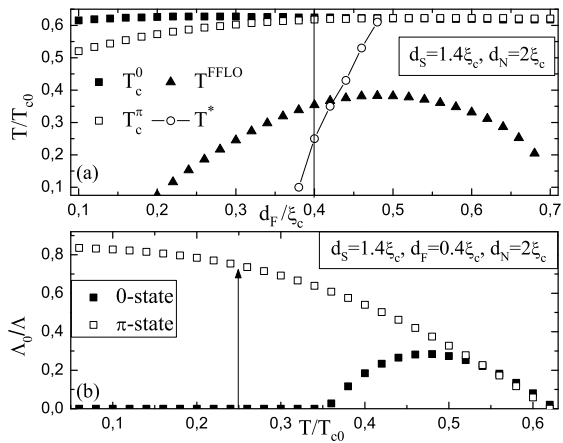}
 \caption{\label{Fig:tc(df)}
(a) Dependence of the critical temperature of 0 ($T_c^0$), $\pi$
($T_c^{\pi})$ and FFLO ($T^{FFLO}$) states on the thickness of F
layer for the S/F/N/F/S pentalayer. Below temperature $T^{*}$ the
$\pi$-state is energetically more favorable than the 0-state. (b)
Temperature dependence of $\Lambda^{-1}$  for the pentalayer with
$d_F=0.4\xi_c$ being in 0 or $\pi$-states. The arrow indicates
temperature of the 0-$\pi$ transition. We use the following
parameters: $h=5k_BT_{c0}$, $d_S=1.4\xi_c$, $d_N=2\xi_c$.}
 \end{center}
\end{figure}

The response of trilayer and pentalayer on the parallel magnetic
field is somewhere different due to different orbital effect
produced by $H$. In trilayer magnetic field induced supervelocity
is maximal in N layer while in pentalayer it is maximal in S
layers. It is the reason why for pentalayer we did not find
enhancement of $T_c$ by parallel magnetic field (shown in Fig.
\ref{Fig:tc(h)} for trilayer) and dependence $\Lambda^{-1}(H)$
like one shown in Fig. \ref{Fig:Lambda}(c). Besides, due to
symmetry of considered pentalayer the parallel magnetic field does
not remove degeneracy of FFLO state with respect to sign of $q_0$
as it does for trilayer. Despite these differences we find that in
FFLO state and at parameters close to FFLO phase domain
$\Lambda^{-1}$ increases in weak magnetic field and decreases in
large field which leads to maximum in dependence $\Lambda^{-1}(H)$
like in trilayer (see Fig.\ref{Fig:Lambda}(a,b)). The only
quantitative difference is that in pentalayer the FFLO state
exists in narrower range of $d_F$ than in trilayer (with the same
parameters) because of its competition with $\pi$ state. Further
in this section we mainly focus on temperature and
current/magnetic field driven $\pi \to $ FFLO transition in
symmetric pentalayer.

Fig.\ref{Fig:tc(df)}(a) demonstrates the dependence of the
critical temperatures of 0 (uniform and FFLO) and $\pi$ (uniform)
states on thickness of F layers. The temperature dependence of
$\Lambda^{-1}$ for the pentalayer being in the FFLO state
resembles that dependence the S/F/N structure being in FFLO state
(compare with Fig. 3(a) from \onlinecite{PRL-2018}). In the
$\pi$-state {\it the same} pentalayer shows monotonic increase of
$\Lambda^{-1}$ with lowering temperature which is typical for
single layer of hybrid S/F or S/F/N structures with no or
negligible contribution of odd frequency triplet superconductivity
to $\Lambda^{-1}$. Similarly to the temperature-driven 0-$\pi$
transition in the S/F/S structures, \cite{PRL-2001-pi,PRB-2014}
there is such a transition in our pentalayer at $T=T^*$ in some
range of $d_F$. In contrast to S/F/S structure considered in Ref.
\onlinecite{PRB-2014} in our pentalayer $\Lambda^{-1}$ increases
at $0 \to \pi$ transition since in the $\pi$-state there is
practically no negative contribution from the triplet component to
$\Lambda^{-1}$. This difference becomes even more dramatic at
transition from 0 FFLO state to uniform $\pi$ state when
$\Lambda^{-1}$ changes from zero up to finite value as temperature
decreases - see Fig. \ref{Fig:tc(df)}(b).

\begin{figure}[hbt]
 \begin{center}
\includegraphics[width=1.0\linewidth]{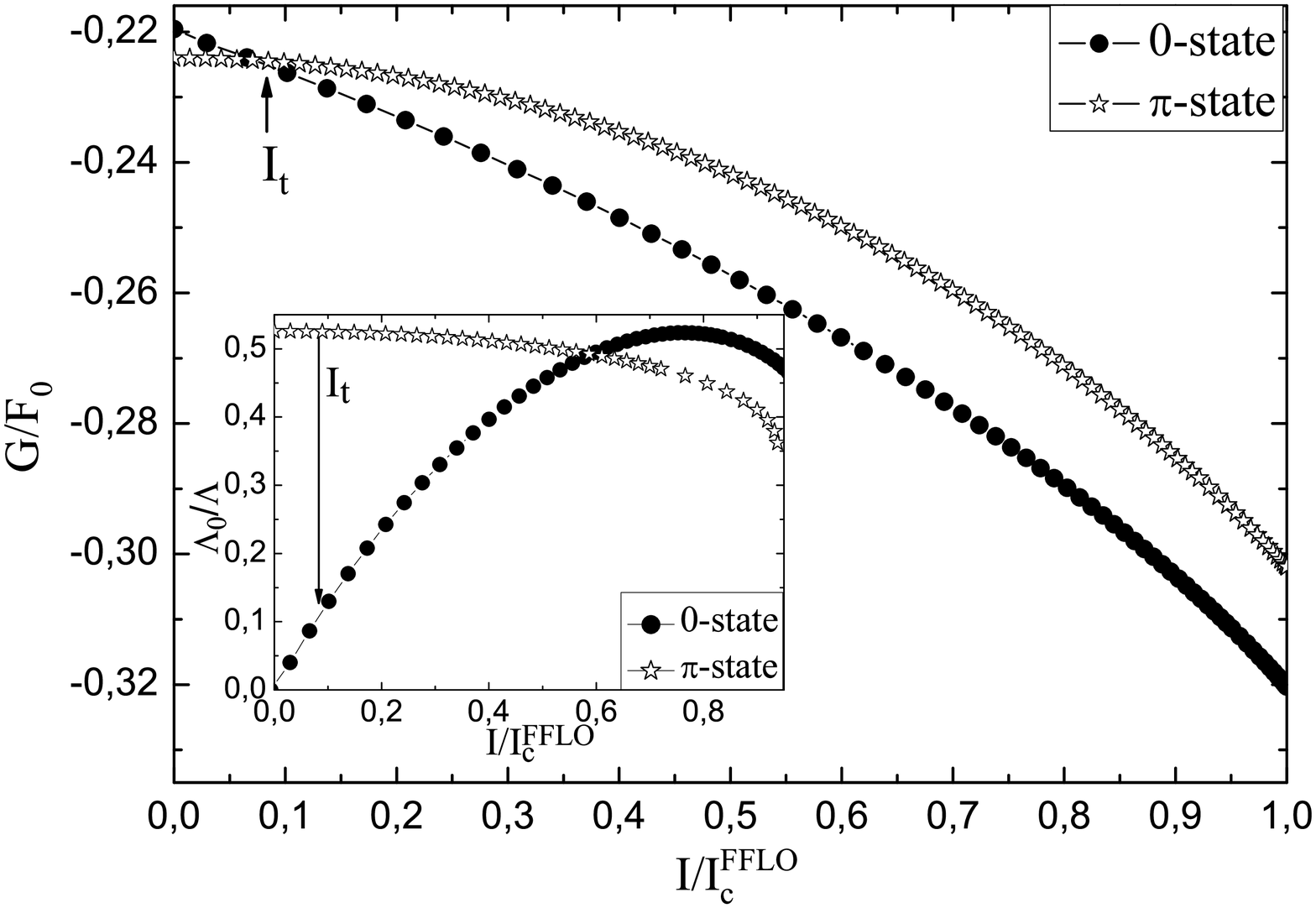}
 \caption{\label{Fig:g(j)}
Current dependence of the Gibbs energy for  0- and $\pi$-states in
the S/F/N/F/S pentalayer at temperature $T=0.2T_{c0}$. Current is
expressed in units of the critical current of the FFLO-state
$I_c^{FFLO}$. The temperature of the 0-$\pi$ transition
$T^{*}=0.26T_{c0}$ at chosen parameters ($h=5k_BT_{c0}$,
$d_S=1.2\xi_c$, $d_F=0.4\xi_c$, $d_N=2\xi_c$). The arrow indicates
the transition current $I_t$ when Gibbs energies of $\pi$ and FFLO
states becomes equal. In the inset we show dependence of
$\Lambda^{-1}$ on supercurrent in both states.}
 \end{center}
\end{figure}

We also find, at fixed temperature $T<T^*<T^{FFLO}$, current or
magnetic driven transition to FFLO state. Let us first consider
current-driven transition. In Fig. \ref{Fig:g(j)} we show
dependence of Gibbs energy $G=F_H-(\hbar/2|e|)Iq_0$, which should
be used for current driven state instead of Helmholtz free
energy,\cite{PR-1968} on current for $\pi$ and FFLO states. One
can see that at $I>I_t$ FFLO state becomes more energetically
favorable. Like as for temperature-driven transition there is a
jump in $\Lambda^{-1}$ (see inset in Fig. \ref{Fig:g(j)}) and in
$q$, because transition current $I_t \sim q\Lambda^{-1}$ is the
same in both states. It implies that $\pi \to$ FFLO transition at
$I>I_t$ should be accompanied by appearance of transitional
electric field, which accelerate superconducting condensate, and
the voltage pulse.

Our calculations show that near $T^{*}$ the transition occurs at
sufficiently small currents $I_t\ll I_c^{FFLO}$, where
$I_c^{FFLO}$ is the critical current of FFLO state (it corresponds
to maximal possible superconducting current flowing along
pentalayer in FFLO state), which is close to $I_c^{\pi}$ of $\pi$
state. With decreasing temperature $I_t$ increases but stays
smaller than $I_c$ for both FFLO and $\pi$ states which makes
current-driven transition possible at all temperatures $0<T<T^*$.
Note that there is also transition from $\pi$ to 0 uniform state
when $T^*>T^{FFLO}$ but it requires larger currents and exists in
narrow temperature interval below $T^*$.

\begin{figure}[hbt]
 \begin{center}
\includegraphics[width=1.0\linewidth]{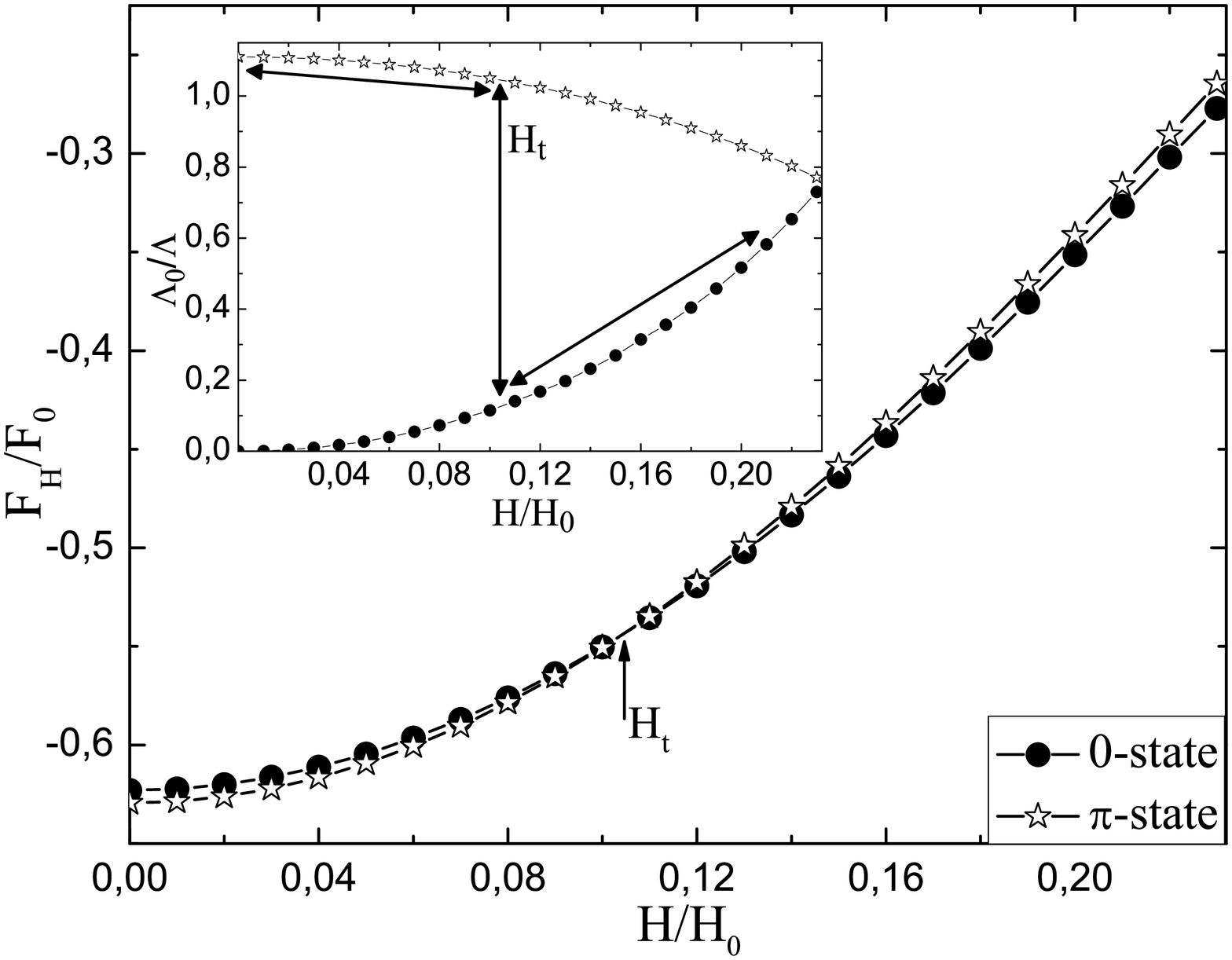}
 \caption{\label{Fig:field}
Dependence of the free energy of the S/F/N/F/S pentalayer on
parallel magnetic field $H$ in the FFLO and $\pi$-states at
temperature $T=0.2T_{c0}$. In the inset we show the dependence of
$\Lambda^{-1}$ on parallel magnetic field $H$ in $\pi$ and FFLO
states. The arrow indicates the field of the $\pi \to $ FFLO
transition. Parameters are the same as in Fig.
\ref{Fig:tc(df)}(b).}
 \end{center}
\end{figure}

The parallel magnetic field differently affects the
superconductivity in the FFLO and $\pi$-states, which at
temperatures $T<T^{*}<T^{FFLO}$ can result to the field-driven
$\pi \to$ FFLO transition (Fig. \ref{Fig:field}). Similar to
current-driven transition here we also have jump in $\Lambda^{-1}$
-- see inset in Fig. \ref{Fig:field} while dependence
$\Lambda^{-1}(H)$ in FFLO state resembles one for the S/F/N
trilayer (compare with Fig. 3(a)). Because energies of $\pi$ and
FFLO states are rather close one needs relatively large magnetic
field to make FFLO state more energetically favorable -- see
Fig.\ref{Fig:field}. Unlike the trilayer, in S/F/N/F/S pentalayer
at certain field $H_{c1}$ vortices can emerge. Using the
expression, valid for the S film with thickness $d_S$,
$H_{c1}\sim\Phi_0/d_S^2$ [\onlinecite{JETP-1970}] and replacing
thickness $d_S$ by total thickness of the pentalayer, we obtain
that $H_{c1}\simeq 0.2 H_0$ for the used in Fig. \ref{Fig:field}
parameters. This estimation explains our choice of maximal
magnetic field in Fig. \ref{Fig:field}. To study effect of
vortices one needs solution of 3D problem and it is out of scope
of our paper.

\section{Summary}

We have studied effect of parallel magnetic field and in-plane
current on screening properties of thin S/F/N and S/F/N/F/S
structures being in or close to FFLO state. In the parameter
region corresponding to the formation of in-plane FFLO-phase, the
effective inverse magnetic field penetration depth $\Lambda^{-1}$
is positive at any finite magnetic field/current and
$\Lambda^{-1}\to 0$ as $H,I\to 0$ which implies diamagnetic
response of such structures. Due to suppression of triplet
superconductivity in F/N layers by magnetic field/current the
dependence $\Lambda^{-1}(H)/(I)$ has unusual field/current
dependence not only in FFLO state but also at parameters close to
FFLO phase domain -- $\Lambda^{-1}$ increases in weak
fields/currents and reaches maximal value at finite $H$/$I$. We
also find that the parallel magnetic field not only control
screening properties of FFLO state but it also can drive S/F/N/F/S
pentalayer from uniform $\pi$ state to in-plane FFLO state which
is accompanied by giant change of $\Lambda^{-1}$. The same
transition could be induced by in-plane current or by changing the
temperature.

Experimentally, predicted effects could be verified, for example
by two-coil technique
\cite{Vodolazov_SUST,JAP-1996,Claassen,Lemberger} which allows to
measure $\Lambda^{-1}$ of thin superconducting structures
directly. Potentially found results could be used in the magnetic
field sensors (due to strong magnetic field dependence of
$\Lambda^{-1}$) or in kinetic inductance detectors of
electromagnetic radiation or particles \cite{Day} when local
heating of heterostructure due to absorbed energy may considerably
change $\Lambda^{-1}$ (see for example Fig. \ref{Fig:tc(df)}),
which determines the kinetic inductance of the sample.

\begin{acknowledgments}
The work was supported by the Russian Science Foundation, grant no
15-12-10020 (D.Yu.V.), in the part of the SFN trilayers and by the
Foundation for the Advancement of Theoretical Physics and
Mathematics BASIS, grant 18-1-2-64-2 (P.M.), in the part of SFNFS
pentalayers.
\end{acknowledgments}

\end{document}